\documentclass{osa-article}

\journal{osac}

\begin{document}

\title{Experimental realization of a high precision tunable hexagonal optical lattice}

\author{Jin-Yu Liu\authormark{1,2,3,$\dagger$}, Guang-Quan Luo\authormark{2,3,$\dagger$}, Xiao-Qiong Wang\authormark{2,3,*}, Andreas Hemmerich\authormark{4}, and Zhi-Fang Xu\authormark{2,3,5,6,*}}

\address{\authormark{1}Department of Physics, Harbin Institute of Technology, Harbin 150001, China\\
\authormark{2}Department of Physics, Southern University of Science and Technology, Shenzhen 518055, China\\
\authormark{3}Shenzhen Institute for Quantum Science and Engineering, Southern University of Science and Technology, Shenzhen 518055, China\\
\authormark{4}Institute of Laser Physics, University of Hamburg, Luruper Chaussee 149, 22761 Hamburg, Germany\\
\authormark{5}International Quantum Academy, Shenzhen 518048, China\\
\authormark{6}Guangdong Provincial Key Laboratory of Quantum Science and Engineering,
Southern University of Science and Technology, Shenzhen 518055, China\\
\authormark{$\dagger$}These authors contributed equally to this work.}

\email{\authormark{*}wangxq@sustech.edu.cn, xuzf@sustech.edu.cn} 

\begin{abstract}
Hexagonal optical lattices offer a tunable platform to study exotic orbital physics in solid state materials. Here, we present a versatile high-precision scheme to implement a hexagonal optical lattice potential, which is engineered by overlapping two independent triangular optical sublattices generated by laser beams with slightly different wavelengths around 1064 nm. This enables us to precisely control the detailed structure of the hexagonal lattice by adjusting the relative position and the relative lattice depth of the two triangular optical sublattices. Taking advantage of the sensitive dependence of the second Bloch band on small lattice deformations, we propose a strategy to optimize the optical lattice geometry with an extremely high precision. This method can also be extended to other lattice configurations involving more than two sublattices. Our work provides the experimental requirements in the search for novel orbital physics of ultracold atoms, for example, in the flat $p$-band of the hexagonal optical lattice.
\end{abstract}

\section{Introduction}
As an additional degree of freedom, different from charge and spin, orbital physics plays a central role for superconductivity, magnetism and transport in transition-metal-oxides due to two fundamental features, i.e., orbital degeneracy and orbital anisotropy\cite{tokura2000orbital}. In periodic crystals, orbitals describe the motion of electrons localized at a lattice site. Electron clouds with different symmetries can arise such as $s$, $p$, and $d$ orbitals, distinguished by their angular momentum. To simulate orbital degrees of freedom with ultracold atoms, one may apply artificial periodic potentials, called optical lattices, by interfering several laser beams, and prepare the atoms in higher Bloch bands. Wannier functions associated with the Bloch bands describe the character of the orbitals. As a versatile platform for quantum simulation~\cite{Jaksch2005, Lewenstein2007, Bloch2008, Bloch2012, Goldman2016, Gross2017}, ultracold atoms in optical lattices have become an ideal system for exploring orbital physics in electronic condensed matter. In contrast to the widely studied $s$ orbital physics in optical lattices~\cite{Bloch2008, lewenstein2012ultracold, Dutta_2015}, the interplay of orbital degeneracy and spatial anisotropy of higher orbitals can induce more complex many-body scenarios with remarkable new features~\cite{Li_2016}.

Over the past decade, $p$ orbitals in optical lattices have been attracting a great deal of attention theoretically. Related experimental developments have provided the means to verify many of the theoretical predictions. In addition to ultracold fermionic gases taking the role of electrons in solid state materials~\cite{PhysRevLett.99.070401, PhysRevB.77.235107, PhysRevB.77.174431, PhysRevLett.100.200406, PhysRevLett.100.160403, PhysRevLett.101.186807, PhysRevA.83.023615, sun_topological_2012, liu_chiral_2014, PhysRevLett.114.100406, xu2017, PhysRevLett.127.033201}, also bosons, for example loaded into the $p$ band, exhibit exotic many-body quantum phases due to interaction effects~\cite{PhysRevLett.97.190406, wirth2011evidence, soltan-panahi_quantum_2012, PhysRevLett.108.175302, PhysRevLett.111.205302, PhysRevLett.111.215302,_lschl_ger_2013, PhysRevLett.114.115301, Kock_2016, PhysRevLett.121.093401, PhysRevB.99.014504, PhysRevB.103.205144, PhysRevResearch.3.033274}. A pioneering experimental work of Ref.~\cite{wirth2011evidence} successfully realized a $p$-orbital superfluid in a bipartite optical square lattice. In this work, bosons condensed at two quasi-momenta located at the centers of the four boundaries forming the rectangular first Brillouin zone (BZ). The corresponding complex-valued locally chiral $p_x\pm ip_y$ order parameter breaks time-reversal symmetry. Recently, experimental efforts on $p$-orbital physics have been directed to hexagonal optical lattices~\cite{weinberg2016breaking, jin2021evidence, wang_evidence_2021, klafka2021bose}. In our recent work \cite{wang_evidence_2021}, we observed the formation of a globally chiral atomic superfluid, induced by interaction-driven time-reversal symmetry breaking in the second Bloch band of a hexagonal boron nitride (BN) optical lattice. The emergent orbital superfluid supports topological Bogoliubov excitations and edge states.

In experiments with bosons in higher Bloch bands with orbital degrees of freedom, time-reversal symmetry and discrete rotation symmetries must be implemented with high precision in order to realize and exploit orbital degeneracies. In this manuscript, we report a versatile high precision scheme to construct a hexagonal BN optical lattice and propose a strategy to precisely optimize the optical lattice geometry. Unlike other schemes~\cite{soltan2011multi, weinberg2016breaking, duca2015aharonov, jin2021evidence}, we create the hexagonal optical lattice by overlapping two triangular optical lattices, which contribute sublattice sites $\mathcal{A}$ and $\mathcal{B}$ to the resulting BN lattice. Each triangular lattice is generated by intersecting three monochromatic running wave laser beams with out-of-plane linear polarization. The potential depth and spatial position of each sublattice can be adjusted independently in the experiment. We swap the potential depths of the two sublattice wells to load atoms into the desired second band. In this process we added an extra step of reducing the superimposed optical dipole trap to further cool the atoms, which turned out crucial for the realization of long-lived condensates with very low temperature.

Since the energy minima of the second band sensitively depend on the relative positions of the two triangular lattices, small variations of the experimental parameters would cause a significant change in the locations of Bragg resonances in momentum spectra recorded by ballistic time-of-flight (TOF) and subsequent absorption imaging. We thus developed a strategy to optimize the optical lattice geometry. First, we calibrate the positions of Bragg peaks from the momentum spectra, which reflect the corresponding quasi-momenta for the energy minima in the second band. Second, we deduce the current experimental parameters assisted by numerical calculations and tune these parameters to correct the lattice deformation. This strategy has been readily applied in the experiment and we believe that it could be easily generalized to other optical lattice systems.

The remainder of this manuscript is organized as follows: In Sec.~\ref{Theoretical}, we introduce the hexagonal optical lattice potential and analyze its deformation with the variation of experimental parameters. In Sec.~\ref{Experimental}, we describe the protocols for building and controlling the experimental setup. In Sec.~\ref{Optimization}, we show the experimental results of the TOF images of the condensates, and discuss our strategy to optimize the optical lattice potential. A summary and an outlook are given in Sec.~\ref{Summary}.

\section{Theoretical scheme for creating hexagonal optical lattice potential}
\label{Theoretical}
Theoretically, we apply three laser beams intersecting at $120^\circ$ angle with respect to each other in the $xy$-plane to construct the hexagonal lattice potential. The polarizations of all laser beams are parallel to the $z$-axis. Each laser beam comprises two spectral components with slightly different wavelengths $\lambda_1$ and $\lambda_2$ around $1064\,$nm. The interference among each set of spectral components with the same frequency generates a triangular lattice. Therefore, the total hexagonal optical lattice potential is a superposition of two triangular lattices with corresponding lattice sites denoted respectively by $\mathcal{A}$ and $\mathcal{B}$ and can be written as $V_h(\mathbf{r})=V_{\lambda_1}(\mathbf{r})+V_{\lambda_2}(\mathbf{r})$ with
\begin{subequations}
\begin{equation} V_{\lambda_1}(\mathbf{r})=-V_1\left\{3+2\sum _{\left< i,j\right> }\cos\left[\left(\mathbf{k}_{i}-\mathbf{k}_{j}\right)\cdot \mathbf{r}+\left(\theta_{i}-\theta_{j}\right)\right]\right\}, \end{equation}
\begin{equation}V_{\lambda_2}(\mathbf{r})=-V_2\left\{3+2\sum _{\left< i,j\right> }\cos\left[\left(\mathbf{k}_{i}-\mathbf{k}_{j}\right)\cdot \mathbf{r}+\left(\theta'_{i}-\theta'_{j}\right)\right]\right\}.
\end{equation}
\label{potential}
\end{subequations}
Here, $V_{1,2}\ge 0$ as $\lambda_{1,2}$ are far red detuned from the atomic resonance. The summation $\left<i,j\right>$ is limited to $\left<1,2\right>$, $\left<2,3\right>$, $\left<3,1\right>$. The wave vectors are given by $\mathbf{k}_{1}=k_\mathrm{L}(-\sqrt3/2,1/2)$, $\mathbf{k}_{2}=k_\mathrm{L}(\sqrt{3}/2,1/2)$, and $\mathbf{k}_{3}=k_\mathrm{L}\left(0,-1\right)$ with $k_\mathrm{L}=2\pi/\lambda_1$. The phase angles are determined by $\theta_{j}=2\pi\nu_1L_j/c+\theta_{0}$ and $\theta^\prime_{j}=2\pi\nu_2L_j/c+\theta_{0}^\prime$, where $\nu_{1,2}=c/\lambda_{1,2}$ are frequencies and $L_j$ denotes the optical path length of the $j$-th beam from the splitting point to the center of the lattice. $c$ denotes the speed of light. $\theta_{0}$ and $\theta_{0}^{\prime}$ include the phase information for laser beams and are irrelevant to the lattice structure. 

\begin{figure}[htbp]
	\centering
	\includegraphics[width=12cm]{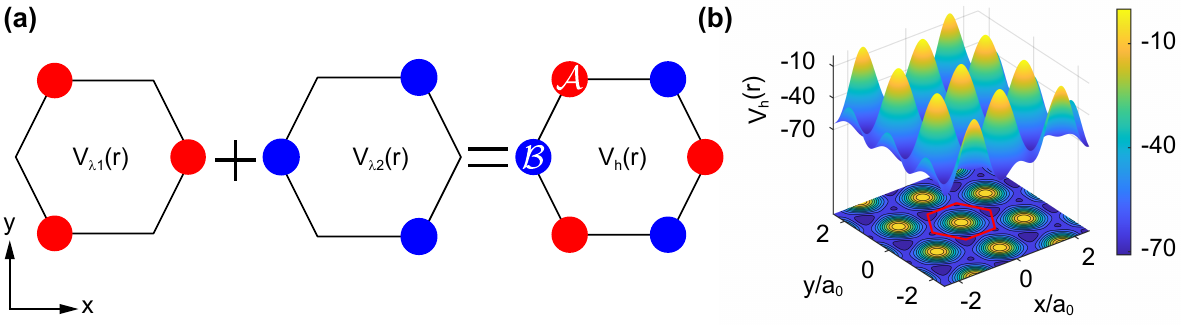}
	\caption{$\mathbf{(a)}$ A hexagonal lattice is obtained by superposing two triangular sublattices. The red and blue filled circles indicate the potential minima of two triangular sublattices. $\mathbf{(b)}$ The hexagonal optical lattice potential with $V_1 = 7.88E_R$ and $V_2 = 7.24E_R$, where $E_R=\hbar^2k_L^2/(2m)$. $k_L=2\pi/\lambda_1$ is the wave numberr of the laser and $m$ is the atom mass of ${}^{87}$Rb. The red hexagon indicates the unit cell.}
	\label{fig1}
\end{figure}

This optical lattice system has various degrees of freedom to be conveniently adjusted in the experiment. $V_{1,2}$ control the potential well depths of the two triangular sublattices independently. The relative position of the two triangular lattices is described by $\Delta\mathbf{r}$ which is defined via the equation $V_{\lambda_1}(\mathbf{r})/V_1=V_{\lambda_2}(\mathbf{r}-\Delta\mathbf{r})/V_2$ and can be explicitly written as
\begin{equation}
	\Delta\mathbf{r}=\frac{\sqrt{3}a_0}{4 \pi}\left(\sqrt{3} \Delta\theta_{12},-\Delta\theta_{12}-2\Delta\theta_{23} \right).
\end{equation}
Here, $a_0=4\pi/(3\sqrt{3}k_\mathrm{L})$ and the relative phase differences $\Delta\theta_{ij}$ are defined as
\begin{equation}
	\Delta\theta_{ij}=(\theta_{i}-\theta_{j})-(\theta_{i}^\prime -\theta_{j}^\prime )=\frac{2\pi\Delta\nu}{c}(L_i-L_j)
	\label{EqPhase}
\end{equation}
with $\Delta\nu=\nu_1-\nu_2$. Note that there are only two independent degrees of freedom as $\Delta\theta_{12}+\Delta\theta_{23}+\Delta\theta_{31}=0$. Thus, $\Delta\mathbf{r}$ can be rewritten as
\begin{equation}
	\Delta\mathbf{r}=\frac{\sqrt{3}a_0\Delta\nu}{2 c}\left(\sqrt{3} L_1- \sqrt{3} L_2,-L_1-L_2+2L_3 \right).
	\label{dr}
\end{equation}
When fixing $L_3=(L_1+L_2)/2$, we can change $L_1-L_2$ or $\Delta\nu$ to tune the relative position of the two triangular sublattices along the $x$-direction. If we further choose $(L_1-L_2, L_2-L_3, L_3-L_1)=(-6.04 ,3.02, 3.02)\ \text{cm}$ and $\Delta \nu= 3.308$ GHz, $\Delta\mathbf{r}=(-a_0,0)$ and a regular hexagonal optical lattice is formed. The unit cell in this case is a hexagon as shown in Fig.~\ref{fig1} (a). Adjacent vertices are associated with potential wells denoted by $\mathcal{A}$ and $\mathcal{B}$, respectively. The potential maximum is located at the center of the unit cell. A typical resulting hexagonal optical lattice potential is shown in Fig.~\ref{fig1} (b).

\section{Experimental construction of hexagonal optical lattices}
\label{Experimental}
\begin{figure}[htbp]
\centering
\includegraphics[width=12cm]{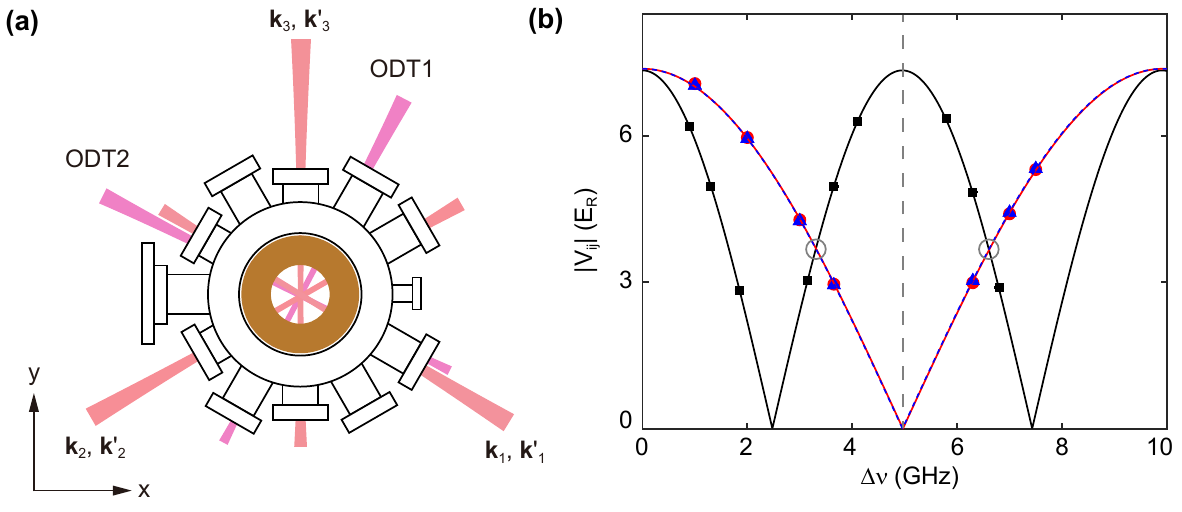}
\caption{$\mathbf{(a)}$ Schematic diagram of the experimental setup. The pancake optical dipole trap is created by the elliptical Gaussian beams colored in purple. The hexagonal lattice is formed by the three Gaussian beams colored in pink. $\mathbf{(b)}$ The lattice depths $(|V_{12}|, |V_{23}|, |V_{31}|)$ are plotted versus the frequency difference $\Delta\nu$. The black squares, red circles, and blue triangles indicate the experimental data of $|V_{12}|$, $|V_{23}|$, and $|V_{31}|$ respectively. The black solid, red solid, and blue dashed lines are the fitting curves for the experimental data symbols with the same color. The central vertical gray dashed line and the gray solid circles mark the values of $\Delta\nu$ for two specific configurations, with a one-dimensional (1D) and hexagonal lattice geometry, respectively.}
\label{fig2}
\end{figure}

Our experiment starts with a $ ^{87}$Rb condensate with approximately $ 4 \times 10^4 $ atoms in the hyperfine ground state $\vert F=1,m_F=-1 \rangle$ without discernable thermal atoms, produced in a pancake optical dipole trap. The dipole trap is formed by two elliptical beams perpendicular to each other at the center of an ultra-high dodecagonal vacuum metal chamber, which is shown in Fig.~\ref{fig2} (a). The trapping frequencies are $\{\omega_x,\omega_y,\omega_z\}=2 \pi \times\{26.4, 26.7, 70.6\}$Hz. A two-dimensional hexagonal optical lattice is formed by superimposing two sets of triangular lattices generated by three two-color beams.

Experimentally, to create a regular hexagonal lattice we need to fine tune all parameters including $L_1$, $L_2$, $L_3$, and $\Delta \nu$. First, we measure and roughly set the values of $(L_1-L_2, L_2-L_3, L_3-L_1)$ as $(-6, 3, 3)$ cm. Second, we turn on two out of three lattice beams to form a 1D optical lattice. Loading a Bose-Einstein condensation into the lattice, we calibrate the lattice depth via amplitude modulated spectroscopy and obtain information for the lattice center difference when we turn on all three lattice beams. As an example, we consider the $i$-th and $j$-th lattice beams with four wave vectors $(\mathbf{k}_{i},\mathbf{k}_{j},\mathbf{k}_{i}^\prime,\mathbf{k}_{j}^\prime)$. In the beginning, we tune the power of each spectral component of each beam to make sure that $V_1=V_2$ (denoted as $V_0$). Their combination gives rise to a 1D lattice potential, which is given by
\begin{eqnarray}
V_{ij}^{\rm (1D)}(\mathbf{r})=-4V_{0}-V_{ij}\cos\left\{\left(\mathbf{k}_{i}-\mathbf{k}_{j} \right)\cdot\mathbf{r}+\frac{1}{2}\left[( \theta_{i} -\theta_{j} )+( \theta_{i}' -\theta_{j}')\right] \right\}
\label{1dlattice}
\end{eqnarray}
where $V_{ij}=4V_{0}\cdot \cos \left(\frac{\pi \Delta v}{c}\left(L_{i}-L_{j}\right)\right)$.

We then change the frequency difference among two spectral components in a range of $(0,10)$ GHz and measure $(|V_{12}|, |V_{23}|, |V_{31}|)$ accordingly. Fig.~\ref{fig2}(b) shows our final results after we carefully set $L_1$, $L_2$, and $L_3$. Note that we use the fact that $V_{23}=V_{31}$ holds for arbitrary values of $\Delta \nu$ when $L_2-L_3=L_3-L_1$ to fine tune $L_3$. In Fig.~\ref{fig2}(b), we fit the experimental data by applying the formula
\begin{eqnarray}
|V_{ij}(\Delta\nu)|=a_{ij}\left|\,\cos\frac{2\pi \Delta\nu}{b_{ij}}\,\right|,
\end{eqnarray}
where $a_{ij}$ and $b_{ij}$ are the fitting parameters. The fitting results are $(a_{12}=7.347(8)\ E_R, b_{12}=9.909(6)\, \rm GHz)$, $(a_{23}=7.38(2)\ E_R, b_{23}=19.87(2)\, \rm GHz)$, $(a_{31}=7.38(1)\ E_R, b_{31}=19.84(2)\, \rm GHz)$ in the confidence level of 95$\%$ for curves.  Based on these parameters, we obtain more precision values for optical path differences as $(L_1-L_2, L_2-L_3, L_3-L_1)=(-6.04 ,3.02, 3.02)$ cm. When the frequency difference is set at $\Delta\nu=3.308\,\rm GHz$ or $6.607\,\rm GHz$ as marked by two circles in Fig.~\ref{fig2}(b), a regular hexagonal is created. 

\section{Optimization of the hexagonal lattice geometry}
\label{Optimization}
\subsection{Bose-Einstein condensation in a hexagonal optical lattice}
To further determine and fine tune the geometry of the hexagonal lattice after turning on all three laser beams, we load atoms into the second Bloch band of a hexagonal BN lattice with $V_1\neq V_2$. Initially, atoms are loaded into the $s$ orbitals in the deep wells of a hexagonal BN lattice. Applying the population swapping technique \cite{0Controlled,folling2007direct} by a rapid quench of the lattice depths in 100 $\mu s$, such that the sizes of $V_1$, $V_2$ and hence the depth of the two subclasses of wells are swapped, the atoms are transferred into the $s$ orbitals of the shallow wells. After a hold time, the atoms recondense at the band mimina of the second Bloch band. The momentum distribution of the condensate is recorded by suddenly switching off all potentials followed by $20\,$ms free expansion before an absorption image is taken. More details can be found in Ref.~\cite{wang_evidence_2021}. We thus distinguish the degree of the deformation of the hexagonal lattice due to improper choices of the optical paths and the frequency difference $\Delta\nu$.

\begin{figure}[htbp]
	\centering
	\includegraphics[width=12cm]{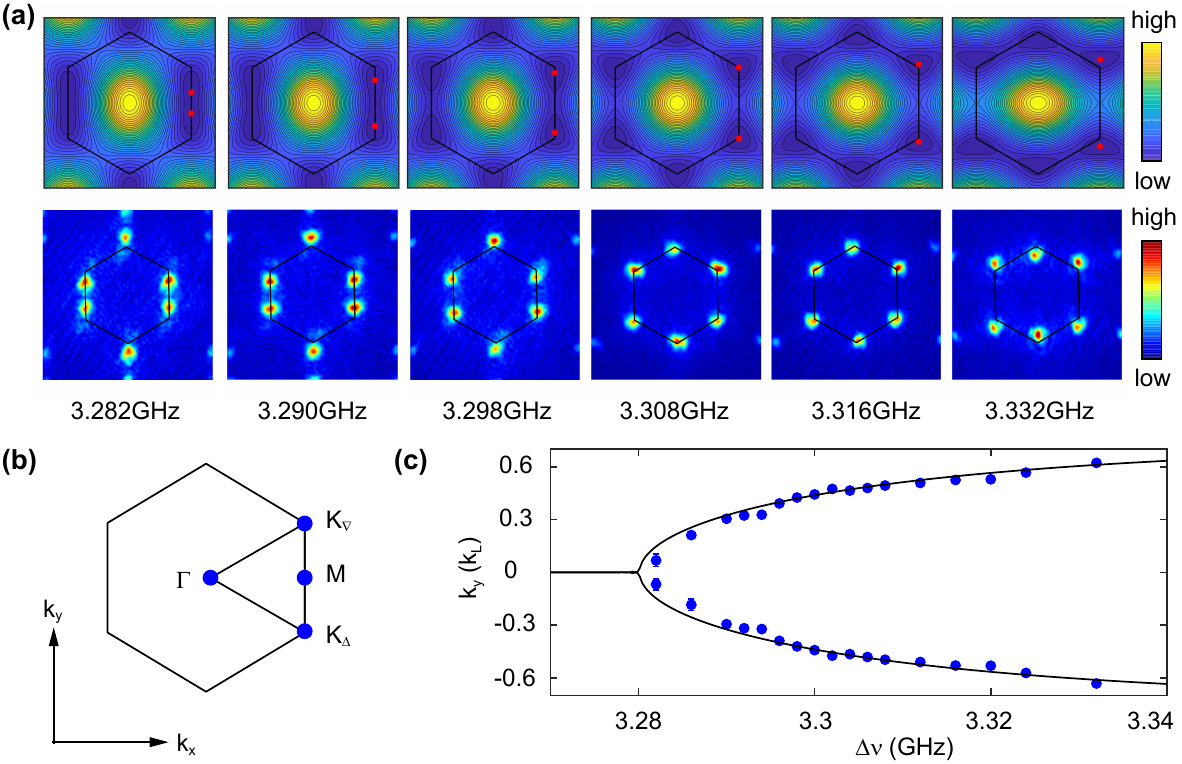}
	\caption{$\mathbf{(a)}$ The upper panel shows the single-particle energy spectra of the second band in the reciprocal space with $(V_1,V_2)=(7.88,7.24)\ E_R$. The red dots indicate the energy minima. The lower panel shows the corresponding experimental TOF images. The black hexagons indicate the first BZ. The frequency differences $\Delta\nu$ are chosen between $3.282\,$GHz to $3.332\,$GHz. $\mathbf{(b)}$ The first BZ of the hexagonal lattice in momentum space. The blue dots indicate the high symmetry points. $\mathbf{(c)}$ The black lines indicate the positions of the energy minima obtained by theoretical calculations. The blue dots indicate the average positions of the atoms obtained from experimental TOF images. The upper and lower blanches indicate the top right corner and bottom right corner, respectively.}
	\label{fig3}
\end{figure}

Fixing the optical paths as $(L_1-L_2, L_2-L_3, L_3-L_1)=(-6.04 ,3.02, 3.02)$ cm, we have adjusted different values of $\Delta\nu$ between $3.282\,$GHz and $3.332\,$GHz. Fig.~\ref{fig3}(a) shows the corresponding theoretically obtained second band dispersions (first row), obtained from plane-wave expansion method, and the experimentally recorded momentum distributions for the condensate in the second band (second row). For the case of $\Delta\nu=3.308\, \rm GHz$, the second band minima are located at the two $K$ points denoted as $K_{\Delta}$ and $K_{\nabla}$ (cf. Fig.~\ref{fig3}(b)). Changing $\Delta\nu$, we observe that the two minima shift along the $k_y$ direction. In Fig.~\ref{fig3}(c), we compare the location of the second band minima with the experimental observed momentum distribution and find that Bragg peaks in experimental TOF images are in good agreement with  theoretical predictions for various $\Delta\nu$. In the low frequency difference limit, two band minima merge together at an $M$ point (cf. Fig.~\ref{fig3}(b)). This phenomenon is quite similar to the band dispersion change in a shaking 1D lattice~\cite{parker2013direct,clark2016universal}. Thus, a similar ferromagnetic quantum phase transition would also occur in our system when we change $\Delta\nu$. 

\subsection{A new optimization strategy}
From the momentum distribution of the condensate in the second band, we find that the positions of the Bragg peaks are very sensitive to the frequency difference $\Delta\nu$ and the optical paths $L_1$, $L_2$, and $L_3$, which does not occur for atoms residing in the lowest band. Hence, we can take advantage of this sensitivity to further fine correct the distortion of the hexagonal optical lattices after implementing the method mentioned in Sec.~\ref{Experimental}. Details of the optimization strategy are presented below.

\begin{figure}[htpb]
	\centering
	\includegraphics[width=12cm]{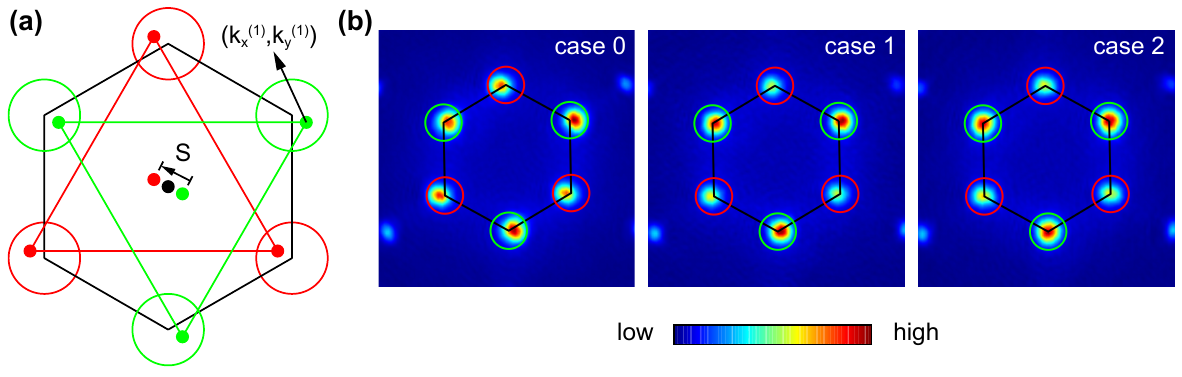}
	\caption{$\mathbf{(a)}$ Sketch of the momentum distribution in a deformed hexagonal lattice. The red and green triangles indicate the three $K_\Delta^\prime$ and $K_\nabla^\prime$ points, respectively. The relative distance between the centers of the two triangles is labeled by $\mathbf{S}$. $\mathbf{(b)}$ The experimental TOF images correspond to $\delta L_3 = -173 \,\mathrm{\mu m}$, $\delta L_3 =-98\, \mathrm{\mu m}$, and $\delta L_3=-20 \,\mathrm{\mu m}$ respectively. The optical lattice potential is set as $(V_1,V_2)=(7.87,7.23)\ E_R$. The black hexagon indicates the first BZ.}
	\label{fig4}
\end{figure}

To avoid confusion, we use $K_\Delta^\prime$ and $K_\nabla^\prime$ to label the locations of the six major Bragg peaks in the momentum distributions in Fig.~\ref{fig4}, which correspond to the condensate momenta in reciprocal space. $K_\Delta$ ($K_\nabla$) and $K_\Delta^\prime$ ($K_\nabla^\prime$) are identical in the case of a perfect regular hexagonal lattice. The three $K_\Delta^\prime$ and $K_\nabla^\prime$ points form two triangles $\Delta^\prime$ and $\nabla^\prime$, respectively. We note that $\Delta^\prime$ and $\nabla^\prime$ are two regular triangles with the same size, which means that  directions and powers of three lattice beams are set accurately in the experiment. In addition, every $K_\Delta^\prime$ point has a inverse $K_\nabla^\prime$ point and vice versa, which is ensured by the time-reversal symmetry of the system. In summary, the only significant error is the misalignment of the two centers of $\Delta^\prime$ and $\nabla^\prime$, which are not identical unless all $K^\prime$ points form a regular hexagon leading to a perfect regular hexagonal lattice (see Fig.~\ref{fig4} (a)).

The displacement from the center of $\nabla^\prime$ to the center of $\Delta^\prime$ is labeled as $\mathbf{S}$. The positions of all $K^\prime$ points can be derived by $\mathbf{S}$. For example, the top right vertex of the hexagon is located at
\begin{equation}
(k_x^{(1)},k_y^{(1)})=(\frac{\sqrt{3}k_L-S(1)}{2},\frac{k_L-S(2)}{2}).
\label{EqS}
\end{equation}
When $\mathbf{S}=0$, we obtain a regular hexagonal lattice with $(k_x^{(1)},k_y^{(1)})=(\sqrt{3}k_L/2,k_L/2)$. When $\mathbf{S}\ne 0$, we only need to optimize the phases $\theta_i$ and $\theta_i^\prime$ by fine tuning $L_3$ and $\Delta\nu$. Due to the constraint $\sum_{\langle i,j\rangle}\Delta\theta_{ij}=0$, we can choose two out of three relative phase differences, e.g. $\Delta\theta_{12}$ and $\Delta\theta_{23}$, to determine a certain displacement $\mathbf{S}$. And a displacement $\mathbf{S}$, in turn, can also be used to derive the correct value of $\Delta\theta_{12}$ and $\Delta\theta_{23}$.

We apply three steps to determine the relative phase differences $\Delta\theta_{12}$ and $\Delta\theta_{23}$. First, we extract the positions of the six major Bragg peaks from a TOF image. The centers of two triangles $\Delta^\prime$ and $\nabla^\prime$ are derived by averaging three $K_\Delta^\prime$ and three $K_\nabla^\prime$ points respectively. The displacement $\mathbf{S}$ between the two triangle centers is thus obtained. Secondly, we use Eq.~\ref{EqS} to calculate the position of the top right vertex $(k_x^{(1)},k_y^{(1)})$, which is more accurate than extracting the position of the corresponding Bragg peak directly. Finally, $\Delta\theta_{12}$ and $\Delta\theta_{23}$ of the TOF image are obtained by finding the solution of $f(\Delta\theta_{12},\Delta\theta_{23})=0$, where the objective function is defined as
\begin{equation}
f(\Delta\theta_{12},\Delta\theta_{23})=\left.\left\{\left[\frac{\partial E_2(\mathbf{k})}{\partial k_x}\right]^2+\left[\frac{\partial E_2(\mathbf{k})}{\partial k_y}\right]^2\right\}\right|_{\mathbf{k}=(k_x^{(1)},k_y^{(1)})}.
\label{Eqf}
\end{equation}
Here, $E_2(\mathbf{k})$ is the energy dispersion for the second band and depends on the values of $\Delta\theta_{12}$ and $\Delta\theta_{23}$. Obviously, $f(\Delta\theta_{12},\Delta\theta_{23})=0$ when $(k_x^{(1)},k_y^{(1)})$ is a minimum of the energy spectrum. We thus optimize the objective function $f(\Delta\theta_{12},\Delta\theta_{23})$ in the neighborhood of $(-4\pi/3,2\pi/3)$ to find the correct values of $\Delta\theta_{12}$ and $\Delta\theta_{23}$. 

Applying this strategy, we can extract $\Delta\theta_{12}$ and $\Delta\theta_{23}$ from an arbitrary TOF image. If their values are different from our target values $\Delta\theta^0_{12}$ and $\Delta\theta^0_{23}$, as estimated we can increase $L_3$ and $\Delta\nu$, respectively, by $\delta L_3$ and $\delta\Delta\nu$, to reach the target values, where 
\begin{equation}
\begin{aligned}
\delta L_{3}=&\left( \frac{\Delta \theta _{23}}{\Delta \theta _{12}}-\frac{\Delta \theta _{23}^{0}}{\Delta \theta _{12}^{0}} \right) \frac{c\Delta \theta _{12}}{2\pi \Delta \nu},\\
\delta\Delta\nu=&\frac{\Delta \theta _{12}^{0}-\Delta \theta _{12}}{\Delta \theta _{12}^{}}\Delta \nu.
\end{aligned}
\end{equation}
When $(\Delta\theta_{12},\Delta\theta_{23})=(-4\pi/3,2\pi/3)$, a regular hexagonal lattice is created. In our experiments, we repeat the measurements and superimpose TOF images to increase the ratio of the signal to noise to evaluate the correct values of $\Delta\theta_{12}$ and $\Delta\theta_{23}$.

To demonstrate the power of the optimization strategy, we perform experiments and tune parameters according to the strategy, which is shown in Fig.~{\ref{fig4}(b). First, we set $\Delta\nu=3.308\, \rm GHz$, and repeat experiments of loading atoms into the second band and hold for $105\,$ms. The corresponding momentum distribution from averaging over 65 experimental runs is shown in the left panel of Fig.~\ref{fig4}(b). After evaluation, we find that to obtain a regular hexagonal lattice, we still need to change $L_3$ and $\Delta\nu$ by small values with $\delta L_3=-173 \mu m$ and $\delta\Delta\nu=1.5\, \rm MHz$. We thus decrease $L_3$ by $80\, \rm \mu m$ and keep frequency difference $\Delta\nu$ the same since the main phase error is from $L_3$. After that, we perform another series of experiment and obtain the momentum distribution shown in the middle panel of Fig.~\ref{fig4}(b). By further evaluation, we obtain $\delta L_3=-98 \, \rm\mu m$ and $\delta\Delta\nu=0.7\, \rm MHz$. Later, we decrease $L_3$ by $80 \, \rm \mu m$ again and perform a third series of experiments to obtain the momentum distribution shown in the right panel of Fig.~\ref{fig4}(b). Finally, we obtain $\delta L_3=-20 \, \rm \mu m$ and $\delta\Delta\nu=0.6\, \rm MHz$. It corresponds to $\Delta\theta_{12}=-4\pi/3+0.0002\pi$ and $\Delta\theta_{23}=2\pi/3-0.0006\pi$, which is close to the precision limit we can achieve. We thus conclude that we have already created a regular hexagonal optical lattice.

\section{Summary and outlook}
\label{Summary}
In conclusion, we apply an effective high-precision scheme to construct a highly tunable precisely hexagonal optical lattice in the experiment. Furthermore, we propose and implement strategies to optimize the lattice geometry by carefully determining the positions of Bragg peaks from momentum distributions of the condensate in the second Bloch band. All experimental results are in good agreement with theoretical calculations. Our work paves the way for investigating many-body physics in the hexagonal optical lattice, especially for novel higher orbital physics, which sensitively depends on the lattice rotational symmetry. The schemes of constructing and optimizing the optical lattice can potentially be extended to other non-square lattices.
\begin{backmatter}
	
\bmsection{Acknowledgments}
This work is supported by the Key-Area Research and Development Program of Guangdong Province (Grant No.~2019B030330001), the National Key R\&D Program of China (Grant No.~2018YFA0307200),  NSFC (Grant No.~U1801661), and a fund from Guangdong province (Grant No.~2019ZT08X324). X.-Q. W. acknowledges support from the China Postdoctoral Science Foundation (No. 2021M691444). A.H. acknowledges support by Cluster of Excellence CUI: Advanced Imaging of Matter of the Deutsche Forschungsgemeinschaft (DFG) - EXC 2056 - project ID 390715994.
	
	\bmsection{Disclosures}
	The authors declare no conflicts of interest.
	
	\bmsection{Data Availability Statement}
	Data underlying the results presented in this paper are not publicly available at this time but may be obtained from the authors upon reasonable request.
	
\end{backmatter}

\end{document}